\documentclass[preprint,aps,prb,superscriptaddress]{revtex4}
\usepackage{multirow}
\usepackage{graphicx}
\usepackage{dcolumn}
\usepackage{bm}

\usepackage{titlesec}
\titleformat*{\section}{\flushleft \bf \large}
\titleformat*{\subsection}{\flushleft \bf}
\titleformat*{\subsubsection}{\flushleft}
\begin{document}

\title{Quantum size effect on the dissociation of O$_2$ molecules on ultrathin Pb(111) films}

\author{Ziyu Hu}
\affiliation{College of Science, Beijing University of Chemical
Technology, Beijing 100029, People's Republic of China}
\affiliation{LCP, Institute of Applied Physics and Computational
Mathematics, P.O. Box 8009, Beijing 100088, People's Republic of
China}
\author{Yu Yang}
\affiliation{LCP, Institute of Applied Physics and Computational
Mathematics, P.O. Box 8009, Beijing 100088, People's Republic of
China}
\author{Bo Sun}
\affiliation{LCP, Institute of Applied Physics and Computational
Mathematics, P.O. Box 8009, Beijing 100088, People's Republic of
China}
\author{Xiaohong Shao}
\affiliation{College of Science, Beijing University of Chemical
Technology, Beijing 100029, People's Republic of China}
\author{Wenchuan Wang}
\affiliation{Laboratory of Molecular and Materials Simulation, Key
Laboratory for Nanomaterials of Ministry of Education, Beijing
University of Chemical Technology, Beijing 100029, People's Republic
of China}
\author{Xucun Ma}
\affiliation{Institute of Physics, Chinese Academy of Sciences,
Beijing 100080, People's Republic of China}
\author{Qikun Xue}
\affiliation{Department of Physics, Tsinghua University, Beijing
100084, People's Republic of China}
\author{Ping Zhang}
\thanks{Corresponding author; zhang\_ping@iapcm.ac.cn (P.Z.).}
\affiliation{LCP, Institute of Applied Physics and Computational
Mathematics, P.O. Box 8009, Beijing 100088, People's Republic of
China}

\begin{abstract}

Using first-principles calculations, we systematically study the
dissociation of O$_2$ molecules on different ultrathin Pb(111)
films. Based on our previous work revealing the molecular adsorption
precursor states for O$_2$, we further explore that why there are
two nearly degenerate adsorption states on Pb(111) ultrathin films,
but no precursor adsorption states exist at all on the Mg(0001) and
Al(111) surfaces. And the reason is concluded to be the different
surface electronic structures. For the O$_2$ dissociation, we
consider both the reaction channels from gas-like and molecularly
adsorbed O$_2$ molecules. We find that the energy barrier for O$_2$
dissociation from the molecular adsorption precursor states is
always smaller than from O$_2$ gases. The most energetically
favorable dissociation process is found to be the same on different
Pb(111) films, and the energy barriers are found to be modulated by
the quantum size effects of Pb(111) films.

\end{abstract}

\maketitle

\section{Introduction}

The adsorption and dissociation of O$_2$ molecules on metal surfaces
are of great importance to the subsequent oxidation reactions and to
the formation of metal oxides \cite{Busnengo}. This is especially
true for the formation of thin oxide films that have been widely
used as catalysts, sensors, dielectrics, and corrosion inhibitors
\cite{Kung}. Thus vast studies have been carried out on the O$_2$
adsorption and dissociation on metal surfaces. During these studies,
the theoretical {\it ab initio} modeling based on the adiabatic
approximation has been proved successful over a wide range for
studying the adsorption and dissociation of O$_2$ on transition
metal surfaces. By calculating the adiabatic potential energy
surface (PES), it has been found that O$_{2}$ molecules will
spontaneously dissociate while adsorbing at reactive transition
metal surfaces like iron (Fe) \cite{Blonski}. For noble transition
metals like gold (Au) \cite{Yotsuhashi}, silver (Ag)
\cite{Nakatsuji}, Copper (Cu) \cite{Alatalo}, platinum (Pt)
\cite{Yotsuhashi,Eichler} and Nickel (Ni) \cite{Eichler}, the
adsorption of O$_{2}$ turns out to depend on the ambient
temperature, and both atomic and molecular adsorptions have been
observed. Remarkably, in all above transition metal systems, the
concept of adiabatic calculation works very well in explaining and
predicting a large amount of physical/chemical phenomena during
dissociation process of O$_{2}$. When the attention is focused on
the nontransition metals with only $sp$ valence electrons, an
uncomfortable gap opens between \textit{ab initio} prediction and
experimental observation. The most notable is the long-term enigma
of low initial sticking probability of thermal O$_{2}$ molecules at
Al(111), which has been measured by many independent experiments
\cite{Ertl,Osterlund} but cannot be reproduced by adiabatic
state-of-the-art density functional theory (DFT) calculations
\cite{Honkala,Your2001,Your2002}. The central problem is that the
adiabatic DFT calculations were unable to find any sizeable barriers
on the adiabatic PES, which has led to speculations that
nonadiabatic effects may play an important role in the oxygen
dissociation process at the Al(111) surface
\cite{Your2001,Kas1974,Kas1979,Kat2004,Wod2004,Hell2003,Hell2005}.
Recently, we present a comparative study on the electronic structure
of an O$_2$ molecule in close to the Be(0001), Mg(0001) and Al(111)
surfaces, and find that the triplet state of O$_2$ is influenced by
the Mg(0001) and Al(111) surfaces, but not by the Be(0001) surface
\cite{Zhang09}. Thus we prove that the adiabatic DFT calculations
are still reliable to study the O$_2$/Be(0001) system, and find out
sizable dissociation energy barriers for O$_2$ molecules on the
Be(0001) surface \cite{Zhang09}. However, in contrast to the
systematical results on the O$_2$ dissociation on transition metal
surfaces, there are still no criteria to judge whether an energy
barrier is needed or whether a precursor molecular-adsorption state
exists for the O$_2$ dissociation on $sp$ metal surfaces. Motivated
by these backgrounds, in this paper we investigate the dissociation
of O$_2$ molecules on another $sp$ metal surface, lead (Pb). We find
that the adiabatic DFT calculation also leads to reliable results
for the O$_2$/Pb(111) system.

Another motive for us to study the O$_2$ dissociation on the Pb(111)
surface is that atomically uniform Pb films with thickness from
several to tens of monolayers (MLs) have been successfully
fabricated on both metal and semiconductor substrates
\cite{Lindgren,Milun}, which can serve as an ideal prototype for
experimental researchers to investigate the surface oxidation of Pb.
However, it is believed that for very thin metal films, the
adsorption and dissociation of O$_2$ molecules becomes more complex
because of the two boundaries. According to the
``particle-in-a-box'' model, electrons confined in a uniform film
are quantized into discrete energy levels along the film-normal
direction, forming the so-called quantum well states. Because the
electron Fermi wavelength of Pb is nearly 4 times the lattice
spacing along the [111] direction \cite{Zhang05}, many properties
like thermal stability, surface energy, and work function of the Pb
films oscillate with thickness with a quasibilayer period
\cite{Upton,Czoschke2,Wei}. Recently, a unique two-step method has
been brought out to efficiently oxidize the the Pb(111) surface
\cite{Ma07,Jiang09}. During this special oxidation process, O$_2$
molecules are firstly introduced at a low temperature and adsorb in
a precursor state, then the Pb(111) surface is oxidized by a
subsequently annealing to room temperature \cite{Jiang09}. Most
interestingly, both the oxygen coverage at low temperature and oxide
coverage after annealing are found to be modulated by the quantum
size effect (QSE) of ultrathin Pb(111) films \cite{Ma07}. These
studies together with other investigations \cite{Aballe04} clearly
prove the modulation of surface oxidation by QSE. To understand the
complicated experimental observations, the systematical
investigations on the adsorption and dissociation of O$_2$ molecules
on ultrathin Pb(111) films are needed.

In our previous work, we have already studied the adsorption
properties for both molecular O$_2$ and atomic O on different
Pb(111) films \cite{Yang08,Sun08}. It is found that O$_2$ molecules
will not spontaneously dissociate on Pb(111) films, and a precursor
molecular adsorption state exists \cite{Yang08}. Here in this work,
we further analyze the electronic interactions for the adsorption of
O$_2$, investigate the dissociation process for O$_2$, and study the
QSE on the dissociation energy barriers. The rest of the paper
organized as follows. In Sec. II, we give details of the
first-principles total energy calculations, which is followed in
Sec. III by our analysis for the adsorption state of O$_2$ on the
Pb(111) surface. In Sec. IV, we study the O$_2$ dissociation on
Pb(111) films, and discuss the QSE on the corresponding energy
barriers. And finally in Sec. VI, we give our conclusions.

\section{Computational methods}

The DFT total energy calculations are carried out using the Vienna
\textit{ab initio} simulation package \cite{VASP}. The generalized
gradient approximation (GGA) of Perdew \textit{et al}. \cite{PW91}
and the projector-augmented-wave (PAW) potentials \cite{PAW} are
employed to describe the exchange-correlation energy and the
electron-ion interaction, respectively. The so-called ``repeated
slab'' geometries \cite{Slab} are employed to model the clean
Pb(111) films ranging from 4 to 8 monolayers (ML). A vacuum layer of
20 \AA~ is adopted to separate the Pb(111) slabs in two adjacent
cells, which is found to be sufficiently convergent from our test
calculations. The O$_2$ was placed on one side of the slab, namely
on the top surface, whereas the bottom layer was fixed. All other Pb
layers as well as the oxygen atoms are free to relax until the
forces on them are less than 0.02 eV/\AA. The plane-wave energy
cutoff is set to 400 eV. A Fermi broadening \cite{FermiBroaden} of
0.1 eV is chosen to smear the occupation of the bands around ${\rm
E}_{F}$ by a finite-$T$ Fermi function and extrapolating to
$T\mathtt{=}0$ K. Integration over the Brillouin zone is done using
the Monkhorst-Pack scheme \cite{Pack} with $13\times13\times1$ grid
points. The calculated lattice constant of bulk Pb is 5.03 \AA, in
good agreement with the experimental value of 4.95 \AA~
\cite{Wyckoff}. The binding energy of spin-polarized O$_{2}$ is
calculated to be $D$=5.78 eV per atom and the O-O bond length is
about 1.235 \AA. These results are typical for well-converged
DFT-GGA calculations. Compared to the experimental \cite{Huber}
values of 5.12 eV and 1.21 \AA~ for O binding energy and bonding
length, the usual DFT-GGA result always introduces an
overestimation, which reflects the theoretical deficiency for
describing the local orbitals of the oxygen. We will consider this
overbinding of O$_{2}$ when drawing any conclusion that may be
affected by its explicit value.

\section{The adsorption properties of O$_2$ on the Pb(111) surface}

The geometries of ultrathin Pb(111) films are firstly relaxed before
our study for the adsorption and dissociation of O$_2$ molecules. It
is found that the surface relaxations display well-defined QSE
\cite{Yang08}. Generally speaking, atoms at the outmost layer tend
to relax inward. In contrast, atoms at the second layer tend to
relax outward. And both the contraction for the first interlayer and
the expansion for the second interlayer exhibit oscillations of
bilayer period, reflecting the effect of the quantum well states
\cite{Wei}.

We then check the reliability of our adiabatic calculations. The
main reason for the wrong conclusions lead by adiabatic
first-principles calculations on the adsorption of O$_2$ on
$sp$-metals lies in that the lowest unoccupied molecular orbital
(LUMO) state of O$_2$ is always aligned with the Fermi level at any
distance between the molecule and metal surfaces such as Al(111),
allowing a partial filling of the empty molecular orbital
\cite{Ciacchi04}. Due to this unphysical (metal-molecule or
intramolecule) charge transfer, the adsorption of O$_2$ on the
Al(111) surface is calculated to be dissociative without any energy
barriers, which contradicts with the experimental observations of
low initial sticking probability at low temperatures
\cite{Ertl,Osterlund}. In our previous work \cite{Zhang09}, we have
shown that the unphysical partial filling of the LUMO does not
happen during the adsorption of O$_2$ on the Be(0001) surface, and
thus we obtained sizeable energy barriers for the dissociation of
O$_2$. Here for the O$_2$/Pb(111) system, molecular adsorption
precursor states exist, so we can not judge the reliability of our
calculations by whether a sizable energy barrier exists or not. We
calculate and show in Fig. 1(a) the electronic density of states
(DOS) of an O$_2$ molecule in close to a 5 ML Pb(111) film. In
comparison with the DOS of O$_2$ in close to the Mg(0001) and
Al(111) surfaces depicted in Figs. 1(b) and (c), we find that the
triplet ground state of the O$_2$ molecule is not influenced at all
in close to the Pb(111) film, but totally disturbed by the Mg(0001)
and Al(111) surfaces due to the unphysical charge transfers.
Therefore, we are sure that the adiabatic DFT calculations will not
lead to unphysical conclusions for the O$_2$/Pb(111) system, as well
as for the O$_2$/Be(0001) system.

To facilitate our discussions, we give some notations at first.
Formally, there are four high-symmetry sites on the Pb(111) surface,
respectively the top (T), bridge (B), hcp (HH), and fcc (FH) hollow
sites. Our previous study has already shown that the high-symmetry
sites play crucial roles in the adsorption O$_2$ molecules on
Pb(111) films \cite{Yang08}. Therefore we do not consider other
surface sites any more. At each surface site, an O$_2$ molecule has
three different high-symmetry orientations, respectively along the x
(i.e., [0$\bar{1}$1]), y (i.e., [$\bar{2}$11]), and z (i.e., [111])
directions. Therefore, we will use T-$x,y,z$, B-$x,y,z$, HH-$x,y,z$,
and FH-$x,y,z$ to respectively represent the 12 high symmetry
adsorption channels for O$_2$.

After geometry optimizations by initially setting the O$_2$ molecule
at 4.0 \AA~ along each adsorption channel, we have got 6 molecular
adsorption states for O$_2$ \cite{Yang08}. By using numbers 1, 2,
and 3 to represent the O$_2$ orientation along the x, y, and z
directions, we name the six adsorption states respectively as HH1,
HH2, HH3, FH1, FH2, and FH3. The atomic geometries for the HH1, HH2
and HH3 states are listed in Figs. 2(b), (c) and (d) respectively.
The geometries for the FH1, FH2 and FH3 states are very similar to
HH1, HH2 and HH3 states, with the adsorbed oxygen atoms at fcc
hollow sites instead of hcp hollow sites. Among the six precursor
states, HH1 and HH2 are the most stable and two nearly degenerate
states. In our previous work, we have already discussed in detail
the interactions between oxygen and Pb for the HH1 state
\cite{Yang08}. Here we further analyze the electronic interactions
for the HH2 state to discuss why the two adsorption states with
different O$_2$ orientations (i.e. HH1 and HH2 states) are almost
degenerate.

Figure 3 shows the calculated difference charge density $\Delta
\rho(\mathbf{r})$ for the HH2 state on the 5 ML Pb(111) film, which
is obtained by subtracting the electron densities of noninteracting
component systems, $\rho^{\rm Pb(111)}(\mathbf{r})+\rho^{\rm
O_2}(\mathbf{r})$, from the density $\rho(\mathbf{r})$ of the HH2
state, while retaining the atomic positions of the component systems
at the same location as in HH2. A positive $\Delta \rho(\mathbf{r})$
here represents charge accumulation, whereas a negative $\Delta
\rho(\mathbf{r})$ represents charge depletion. Figure 3(a) clearly
shows a charge transfer from Pb to O$_2$, which is consistent with
the electronegativity difference between O and Pb atoms, and same
with what we found for the HH1 state \cite{Yang08}. The contour plot
for $\Delta \rho(\mathbf{r})$ in the plane containing the two O
atoms and parallel to the Pb surface is shown Fig. 3(b), we can see
that there is an electron depletion area in the middle of the two
oxygen atoms, indicating the donation of electrons from bonding
orbitals of O$_2$ to Pb. We can also see two electron accumulation
areas around the two oxygen atoms indicating the back donation of
electrons from Pb to the $\pi_p^{\star}$ antibonding orbital of
O$_2$. This interaction mechanism of O$_2$-to-surface donation and
surface-to-O$_2$ $\pi_p^{\star}$ back donation is just the same with
that in the HH1 state \cite{Yang08}. So it is no wonder that the two
adsorption states are nearly degenerate and have similar adsorption
energies.

In comparison with the Be(0001), Mg(0001) and Al(111) surfaces,
Pb(111) is unique in that a parallel O$_2$ molecularly adsorbs only
on it. This is because of the different surface electronic
structures. As shown in Fig. 4, the $p$ electronic states around the
Fermi energy strongly hybridize with $s$ states for both the Mg and
Al surfaces (also for the Be(0001) surface \cite{Zhang09}), but does
not hybridize at all for the Pb(111) surface. As we have seen
\cite{Yang08}, the interaction between O$_2$ and the Pb(111) surface
is mainly contributed by the electronic hybridizations between the
frontier molecular orbitals (the orbitals nearest to the Fermi
energy) of O$_2$ and electronic states of Pb. According to the
frontier orbital theory, the interactions between O$_2$ and other
metals have similar mechanisms. However, the frontier orbitals of
O$_2$ (i.e. $\sigma_p$, $\pi_p$, and $\pi_p^{\star}$) are composed
of $p$ electrons of oxygen, thus they are more compatible with $p$
electronic states than $sp$ hybridized electronic states. This
compatibility between frontier orbitals of O$_2$ and $p$ states of
Pb finally leads to the molecular adsorption precursor states.

\section{Dissociation of O$_2$ molecules}

After the studies and discussions for the adsorption on different
Pb(111) films, we now investigate the O$_2$ dissociation. Because of
the existence of precursor states, there are two different kinds of
dissociation processes for O$_2$, respectively from O$_2$ gases and
from the molecular adsorption states. Because the adiabatic
calculations have been proved reliable for the O$_2$/Pb(111) system,
we will calculate the adiabatic one-dimensional (1D) and
two-dimensional (2D) potential energy surface (PES) cuts to evaluate
the dissociation energy barriers for O$_2$ molecules.

The 1D PES cuts for an O$_2$ molecule along the T-$x,y$ and B-$x,y$
channels are show in Fig. 5, together with the O-O bond length as
functions of the O$_2$ height $h$ from the 5 ML Pb(111) film
surface. The adsorption from O$_2$ gas along the FH- and HH-$x,y$
channels will evolve into the molecular adsorption precursor states.
Clearly, the dissociative adsorption of O$_2$ along the T-$x,y$ and
B-$y$ channels are direct and activated, but no dissociative
adsorption is found for an O$_2$ molecules along the B-$x$ channel.
This is different from the O$_2$/Be(0001) system, where the O$_2$
dissociation along B-$x$ channel also occurs \cite{Zhang09}. Among
the calculated channels, the lowest dissociation energy barrier is
found to be 0.36 eV along the B-$y$ channel, with the O-O bond
length and O$_2$ height at the transition state to be 1.33 and
2.20\AA. The 1D PES cuts along the T-$x,y$ channels have similar
distributions, and the corresponding energy barriers are 0.68 and
0.59 eV.

The electronic interactions are then analyzed for the dissociation
process from O$_2$ gases. Figure 6 shows the PDOS for an O$_2$
molecule at the initial, transition and final states of the
dissociation path along the B-$y$ channel. We can see that at the
initial state, the O$_2$ molecule keeps its triplet electronic
structure. As we have discussed, the LUMO of O$_2$ is above the
Fermi energy, proving the reliability of our adiabatic DFT
calculations. The highest occupied molecular orbital (HOMO) and LUMO
are the spin-up and spin-down antibonding $\pi_p^{\star}$ molecular
orbitals, respectively. Besides of HOMO and LUMO, $p$ electrons of
O$_2$ make up of two bonding orbitals, $\sigma_p$ and $\pi_p$. For
both spins, the $\sigma_p$ orbitals are lower in energy than the
$\pi_p$ orbitals. When the O$_2$ molecule evolves into the
transition state, electrons transfer from the $\sigma_p$ and $\pi_p$
molecular orbitals to Pb through hybridizations. As shown in Fig.
6(b), the $\sigma_p$ and $\pi_p$ peaks are diminished. We can also
see from Fig. 6(b) that the HOMO and LUMO of O$_2$ are broadened
through hybridizations with electronic states of Pb. During the
electronic hybridizations, the spin splittings for all the O$_2$
molecular orbitals are all decreased. In the final dissociated
adsorption state, the total spin of the adsorption system becomes
zero. We can see from Fig. 6(c) that the molecular orbitals of O$_2$
no longer exist after electronic hybridizations. As discussed in our
previous work \cite{Sun08}, the interactions between O and Pb atoms
display a mixed ionic/covalent character.

The dissociation of O$_2$ from the molecular adsorption precursor
states are more complicated than from O$_2$ gases because the
formation and breaking of O-Pb bonds during the dissociation of
O$_2$ also need to be considered. Detailed information on the atomic
adsorption of oxygen on the Pb(111) surface can help us to simplify
the problem. In our previous study \cite{Sun08}, we have found that
the O atoms choose to adsorb at hollow sites on the Pb(111) surface.
Thus we only need to consider the dissociation of an adsorbed O$_2$
molecule into different hollow sites. For an O$_2$ molecule in the
HH1 or HH2 state, there are two different dissociation paths,
respectively to the atomic adsorption of oxygen at two neighboring
fcc and hcp hollow sites and to the atomic adsorption of oxygen at
two fcc hollow sites. The calculated 2D PES cuts for the adsorbed
O$_2$ molecules in HH1 and HH2 states to dissociate into two
neighboring hollow sites are shown in Figs. 7(a) and (b), while the
calculated 2D PES cuts for O$_2$ to dissociate into two fcc hollow
sites are also shown in Figs. 7(c) and (d). The calculated energy
barriers for these four paths are respectively 0.22, 0.26, 0.59 and
0.37 eV. These values indicate that the most energetically favorable
dissociation path for an adsorbed O$_2$ molecule is from the HH1
adsorption state into the atomic adsorption of oxygen at two
neighboring fcc and hcp hollow sites. The saddle point in Fig. 7(a)
corresponds to the adsorption structure of two oxygen atoms near two
neighboring bridge sites, with their horizontal distance to be 1.40
\AA. The smallest energy barriers for an O$_2$ molecule in the FH1
and FH2 states to dissociate are both found to be 0.25 eV, which are
larger than the smallest energy barrier from the HH1 state of 0.22
eV.

The electronic properties are then analyzed for the dissociation
path shown in Fig. 7(a), whose energy barrier is the smallest during
the dissociation from molecular adsorption precursor states. The
PDOS around the two oxygen atoms are calculated for the
corresponding initial, transition and final states. As shown in Fig.
8(a), the $\pi_p^{\star}$ antibonding orbital of O$_2$ are broadened
and shifted below the Fermi energy in the HH1 adsorption state,
through hybridizations with $p$ electronic states of Pb. At the
transition state of the dissociation process, the energy interval
between the $\sigma_s$ and $\sigma_s^\star$ orbitals is decreased
from 6.58 to 6.24 eV, and the oxygen $2p$ states hybridize in a
different way with electronic states of Pb from that in the HH1
adsorption state. At the end of the dissociation process, the two
oxygen atoms relax into two neighboring hollow sites, and the
molecular orbitals of O$_2$ all disappear after electronic
hybridizations.

After the systematic studies on the O$_2$ dissociation in different
ways, we reveal that the most energetically favorable one is from
the HH1 adsorption state into two neighboring hollow sites. It is
always believed that because of the confinement of the two
boundaries in the vertical direction, the quantum behavior of
electrons will cause a number of unique properties for thin metal
films that could be superior to their bulk counterpart. In
particular, quantum size effects on surface reactivity have been
studied extensively for applications in catalysis, corrosion, and
gas sensing. Here we further consider the dissociation path for
O$_2$ molecules on different Pb(111) films, including both the
dissociation along the T- and B-$x,y$ channels from O$_2$ gases, and
the dissociation from molecular adsorption states. After systematic
calculations, we find that the most energetically favorable
dissociation paths are the same on different Pb(111) films, i.e.
from the HH1 adsorption state to the atomic adsorption of two oxygen
atoms at two neighboring hollow sites. Figure 9 shows the 2D PES
cuts for the O$_2$ dissociation on 4$\sim$7 monolayers Pb(111)
films. The calculated energy barriers are respectively 0.26, 0.22,
0.26 and 0.24 eV. Since many properties of Pb(111) films such as
work function and surface energy show bilayer oscillation behaviors
because of QSE \cite{Wei}. The even-odd oscillation of the
dissociation energy barrier proves the modulation of QSE on the
O$_2$ dissociation on Pb(111) films.

\section{ CONCLUSION}

In summary, we have investigated the adsorption and dissociation of
O$_2$ molecules on different Pb(111) films. The electronic states of
an O$_2$ molecule in close to the Pb films are firstly analyzed and
imply that the adiabatic calculations are reliable. For the
molecular adsorption precursor states, we find that the electronic
interactions are very similar in the HH1 and HH2 states, and they
are thus two nearly degenerate states. Besides, we further point out
that no $s$-$p$ hybridizations in Pb(111) films is the main reason
for the existence of molecular adsorption precursor states. Based on
our adiabatic PES calculations, we find that the dissociation of
O$_2$ from the molecular adsorption precursor states have lower
energy barriers than from O$_2$ gases. The most energetically
favorable dissociation paths on different Pb(111) films are found to
be the same, which is from the HH1 adsorption state to the atomic
adsorption of two oxygen atoms at two neighboring hollow sites. The
energy barriers of O$_2$ on different Pb(111) films are found to be
modulated by the QSE of Pb(111) films, and show a bilayer
oscillation behavior. Our study further complete the theoretical
picture for the adsorption and dissociation of O$_2$ molecules on
different Pb(111) films.

\begin{acknowledgments}
This work was supported by the NSFC under grants No. 10604010,
10904004, and 60776063.
\end{acknowledgments}

\clearpage

\noindent\textbf{List of captions} \\

\noindent\textbf{Fig.1}~~~ (color online) The electronic density of
states for an O$_2$ molecule in close to the 5 ML Pb(111) film (a),
the Mg(0001) surface (b), and the Al(111) surface (c). The Fermi
energies are all set to 0. \\

\noindent\textbf{Fig.2}~~~ (color online) (a) The considered twelve
high-symmetry adsorption channels for O$_2$ on the Pb(111) surface.
(b), (c) and (d) The three molecular adsorption states for O$_2$ at
the surface hcp hollow site. Red and grey balls represent O and Pb
atoms. There is a Pb atom in the second layer below the HH site to
make it distinct from the FH site. \\

\noindent\textbf{Fig.3}~~~ (color online) (a) Difference charge
density for the HH2 adsorption state of O$_2$ on the 5 ML Pb film.
Red and gray atoms represent O and surface Pb. Regions of electrons
accumulation/depletion are displayed in blue/yellow, respectively,
and the isosurface value is $\pm$0.02 e/\AA. (b) Contour plots of
the difference charge density in the plane that contains the two
oxygen atoms and is parallel to the Pb(111) surface. \\

\noindent\textbf{Fig.4}~~~ (color online) The projected density of
states for the 5 ML Pb(111) film (a), Mg(0001) (b) and Al(111) (c)
surfaces. The Fermi energies are all set to 0. \\

\noindent\textbf{Fig.5}~~~ (color online) One-dimensional cuts of
the potential energy surfaces and the corresponding O-O bond lengths
($d_{\rm O-O}$) as functions of the O$_2$ distance $h$ from the 5 ML
Pb(111) film for eight different dissociative channels. The inset in
each panel indicates the initial (red circles) and final (green
circles) atomic positions of the two oxygen atoms. \\

\noindent\textbf{Fig.6}~~~ (color online) The projected density of
states around the two oxygen atoms at the initial (a), transition
(b) and final states (c) along the B-$y$ adsorption channel. The
Fermi energies are set to 0. \\

\noindent\textbf{Fig.7}~~~ (color online) Color-filled contour plots
of the potential energy surface for O$_2$ dissociation on the 5 ML
Pb(111) film from HH1 (a) and HH2 (b) states to the atomic
adsorption of two oxygen atoms at neighboring hcp and fcc hollow
sites, and from HH1 (c) and HH2 (d) states to the atomic adsorption
at two fcc hollow sites, as functions of the O$_2$ bond length
$d_{\rm O-O}$ and distance $h$ (from the surface). \\

\noindent\textbf{Fig.8}~~~ (color online) The projected density of
states around the two oxygen atoms at the initial (a), transition
(b) and final states (c) of the dissociation path from the HH1 state
to the atomic adsorption of two oxygen atoms at neighboring hcp and
fcc hollow sites. The Fermi energies are set to 0. \\

\noindent\textbf{Fig.9}~~~ (color online) Color-filled contour plots
of the potential energy surface for O$_2$ dissociation on the 4 (a),
5 (b), 6 (c) and 7 ML (d) Pb(111) films from the HH1 state to the
atomic adsorption of two oxygen atoms at neighboring hcp and fcc
hollow sites, as functions of the O$_2$ bond length $d_{\rm O-O}$
and distance $h$ (from the surface). \\

\clearpage

\begin{figure}
\includegraphics[width=1.0\textwidth]{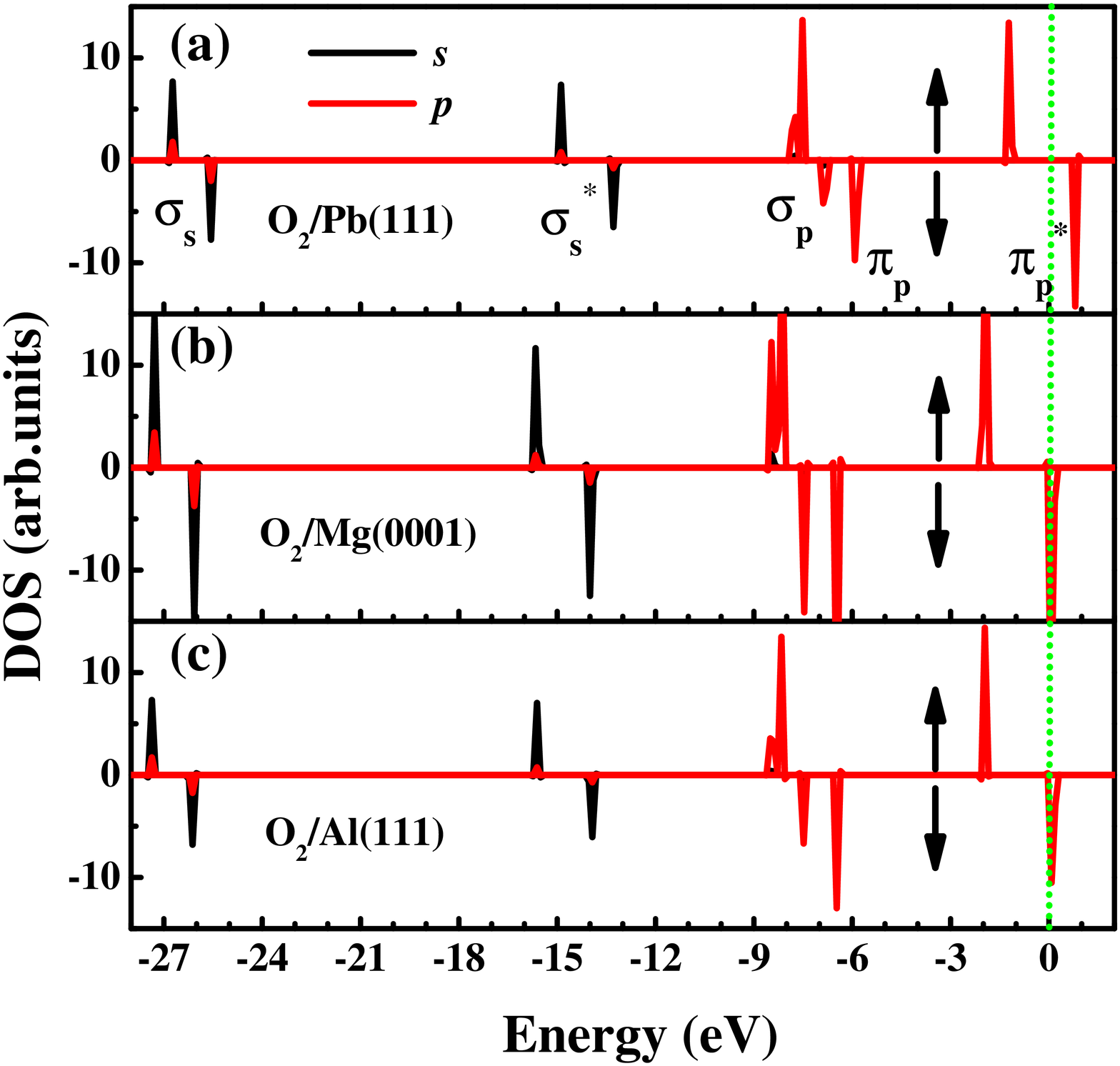}
\caption{\label{fig:fig1}}
\end{figure}
\clearpage
\begin{figure}
\includegraphics[width=1.0\textwidth]{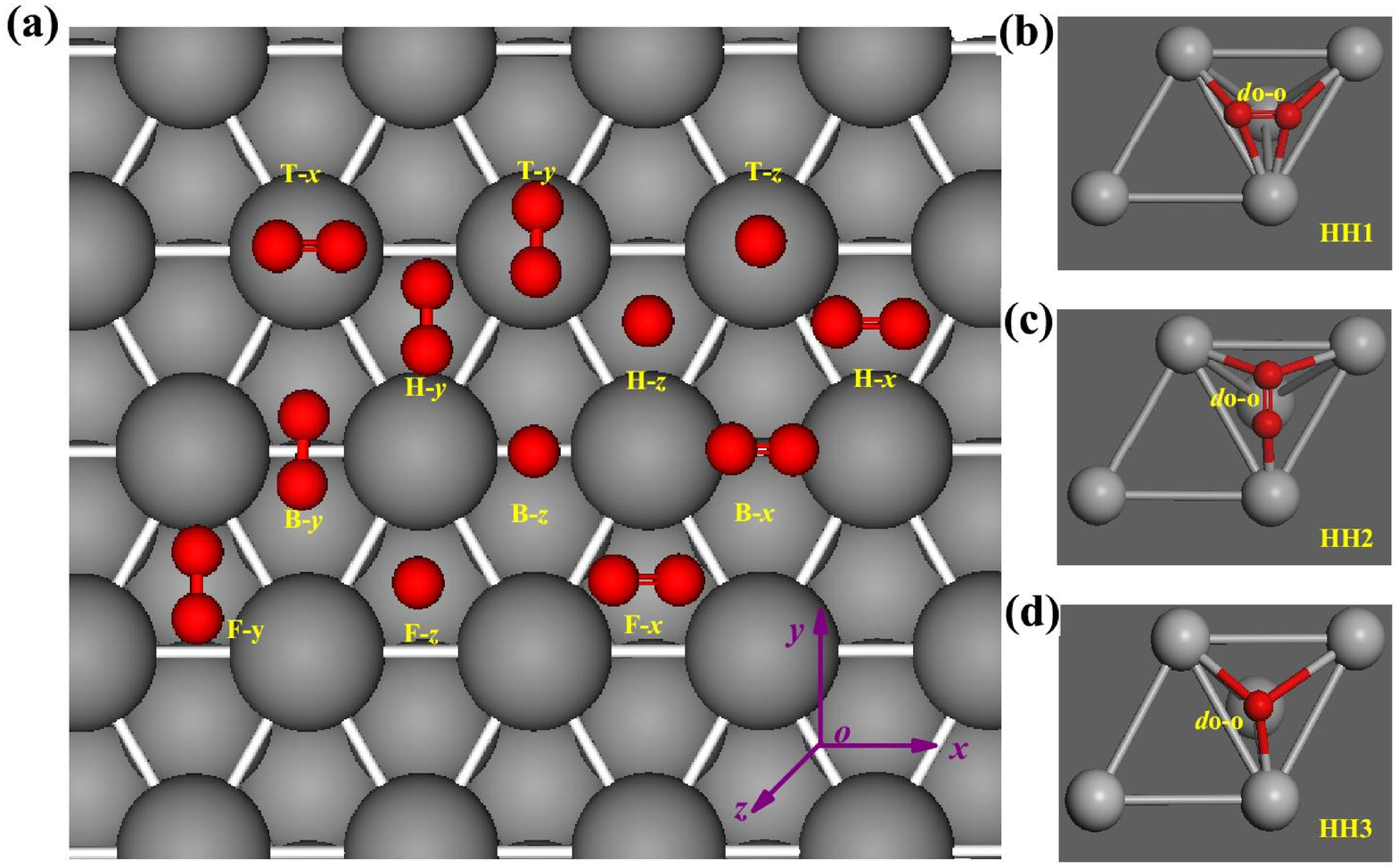}
\caption{\label{fig:fig2}}
\end{figure}
\clearpage
\begin{figure}
\includegraphics[width=1.0\textwidth]{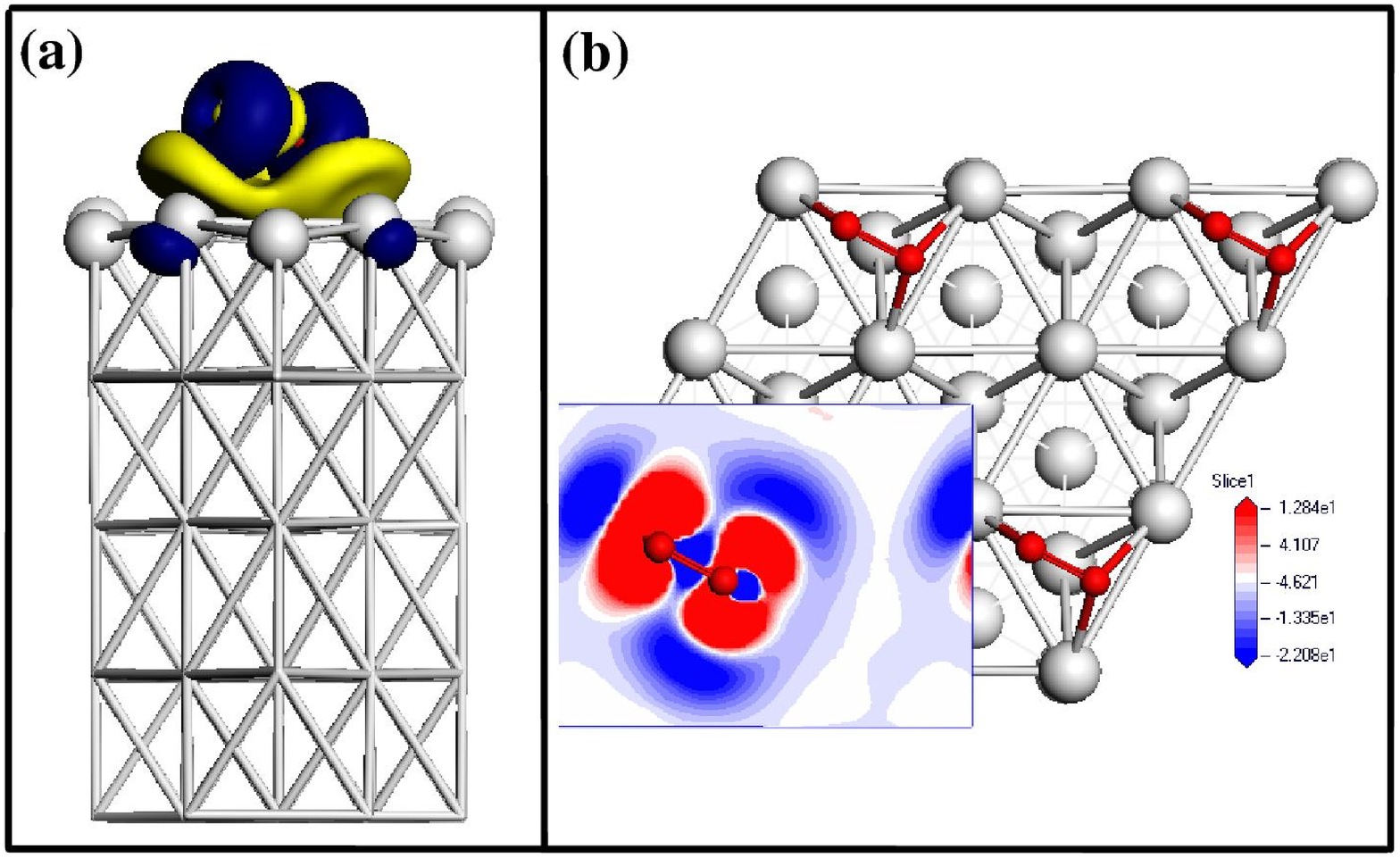}
\caption{\label{fig:fig3}}
\end{figure}
\clearpage
\begin{figure}
\includegraphics[width=1.0\textwidth]{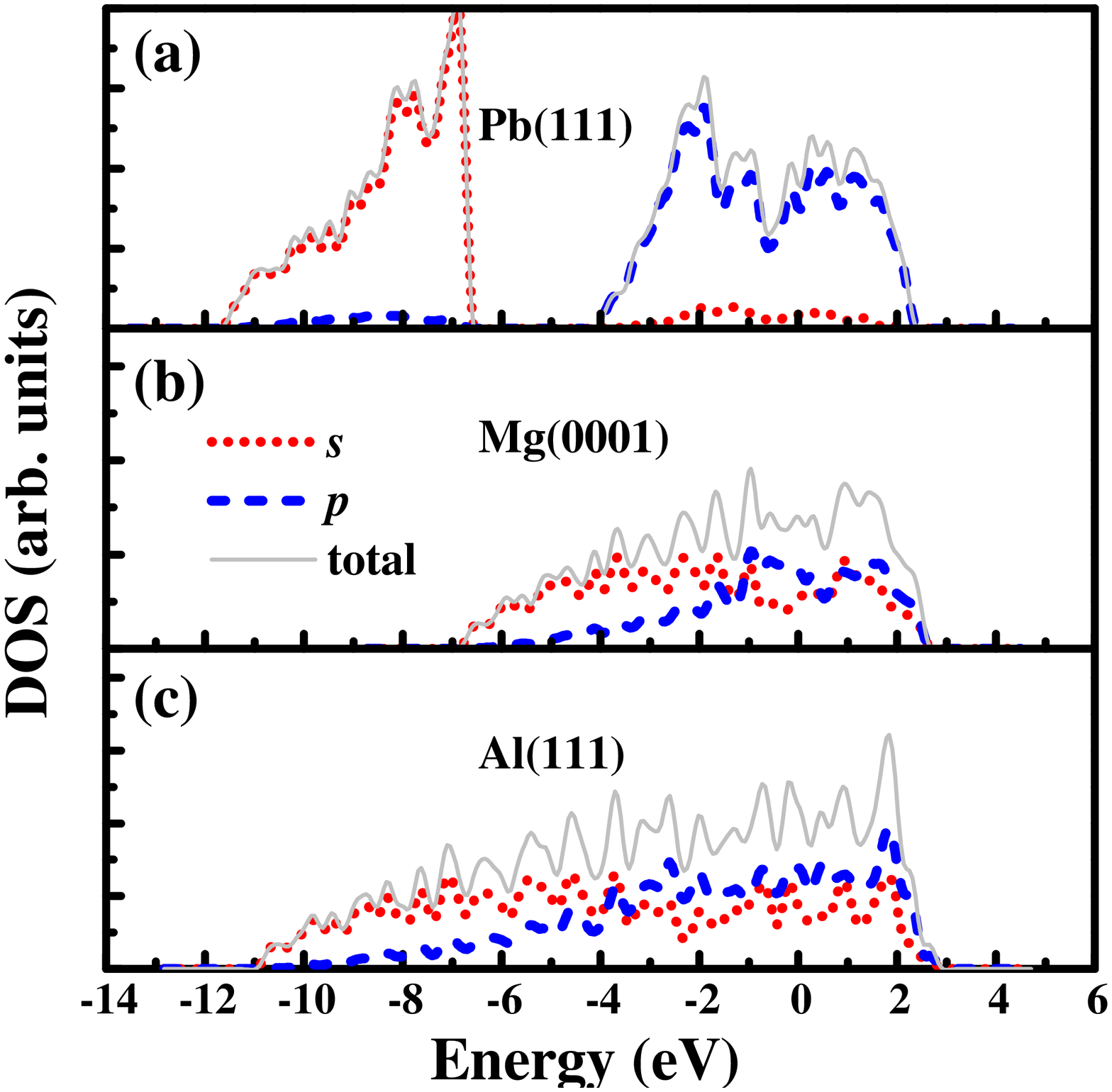}
\caption{\label{fig:fig4}}
\end{figure}
\clearpage
\begin{figure}
\includegraphics[width=1.0\textwidth]{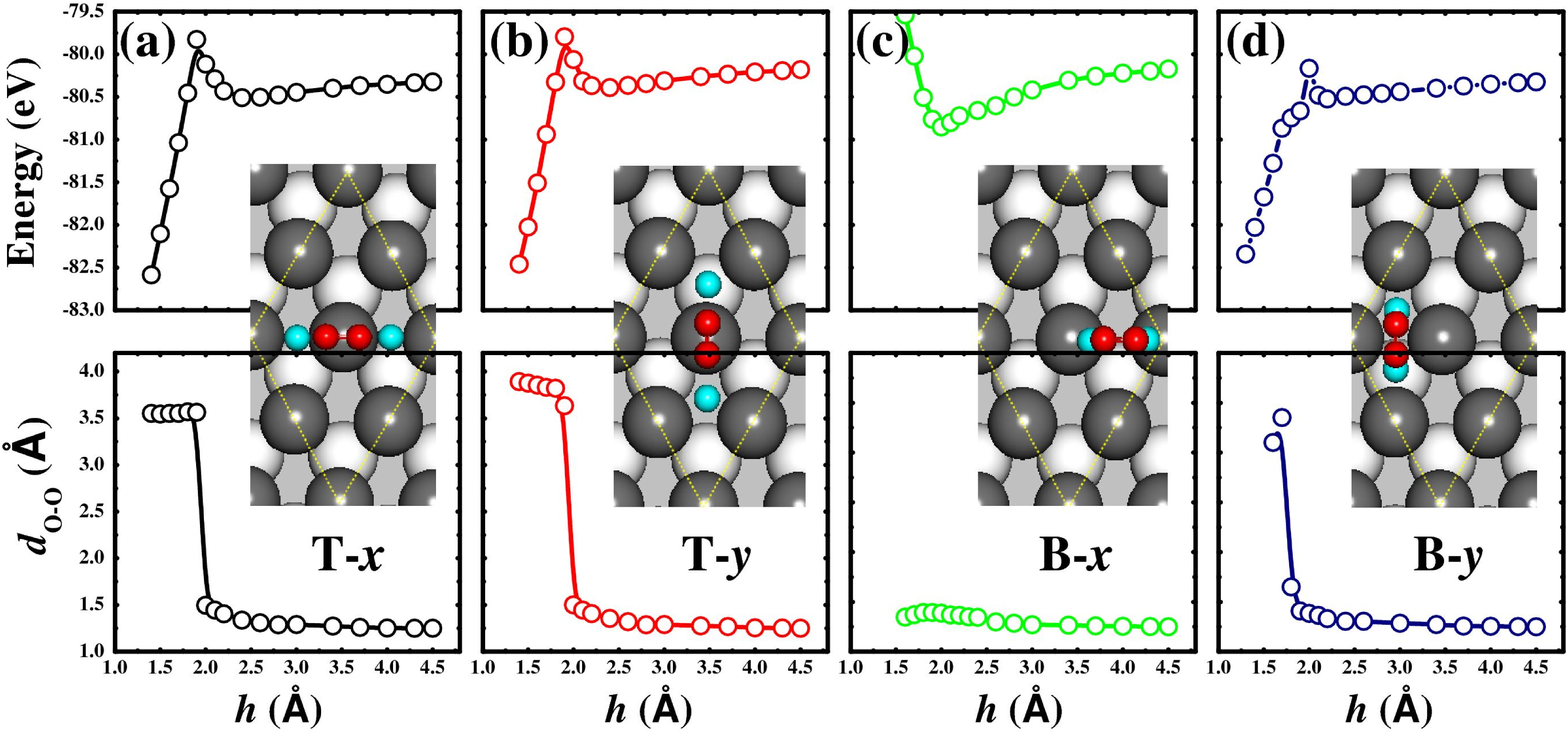}
\caption{\label{fig:fig5}}
\end{figure}
\clearpage
\begin{figure}
\includegraphics[width=1.0\textwidth]{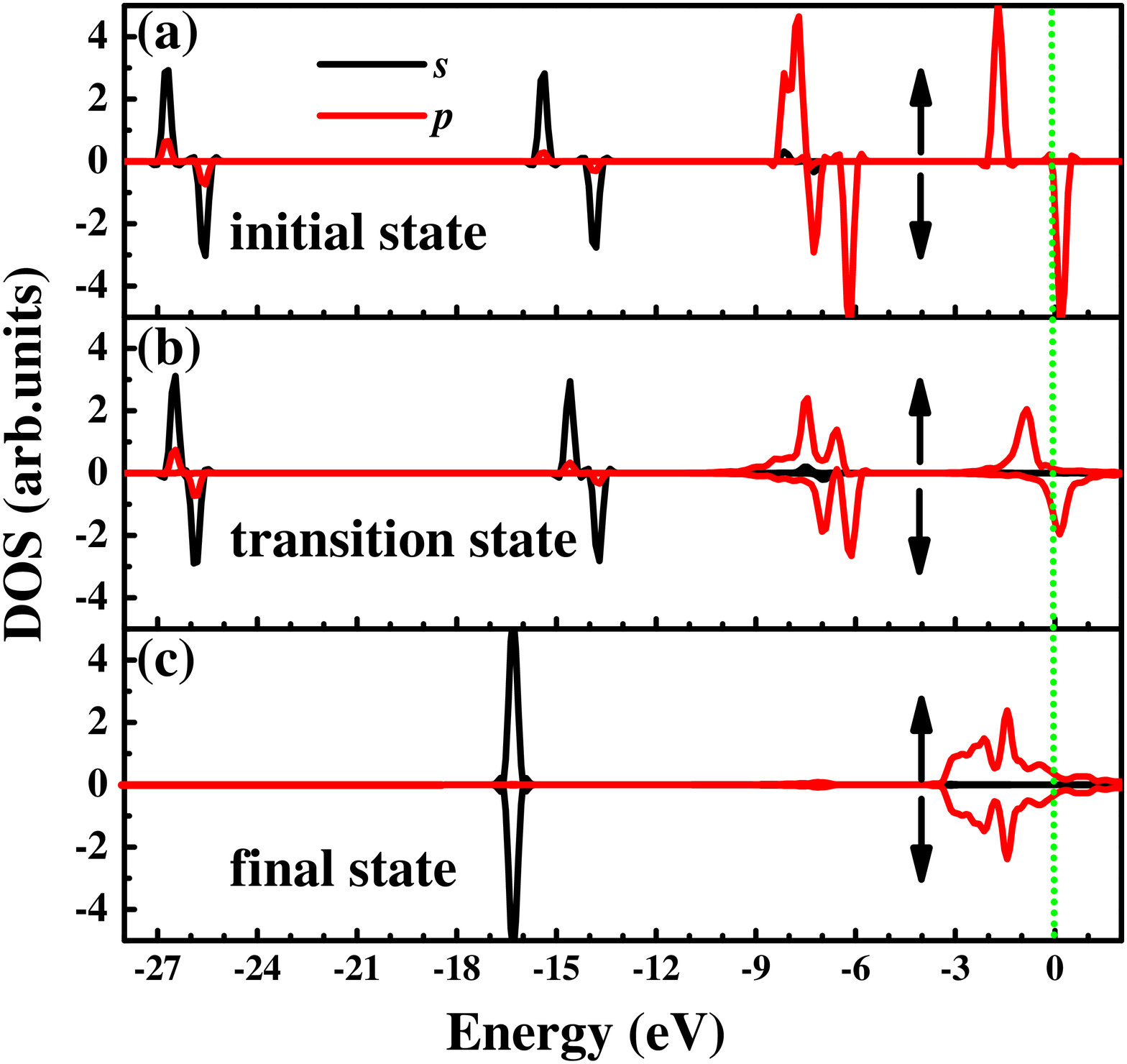}
\caption{\label{fig:fig6}}
\end{figure}
\clearpage
\begin{figure}
\includegraphics[width=1.0\textwidth]{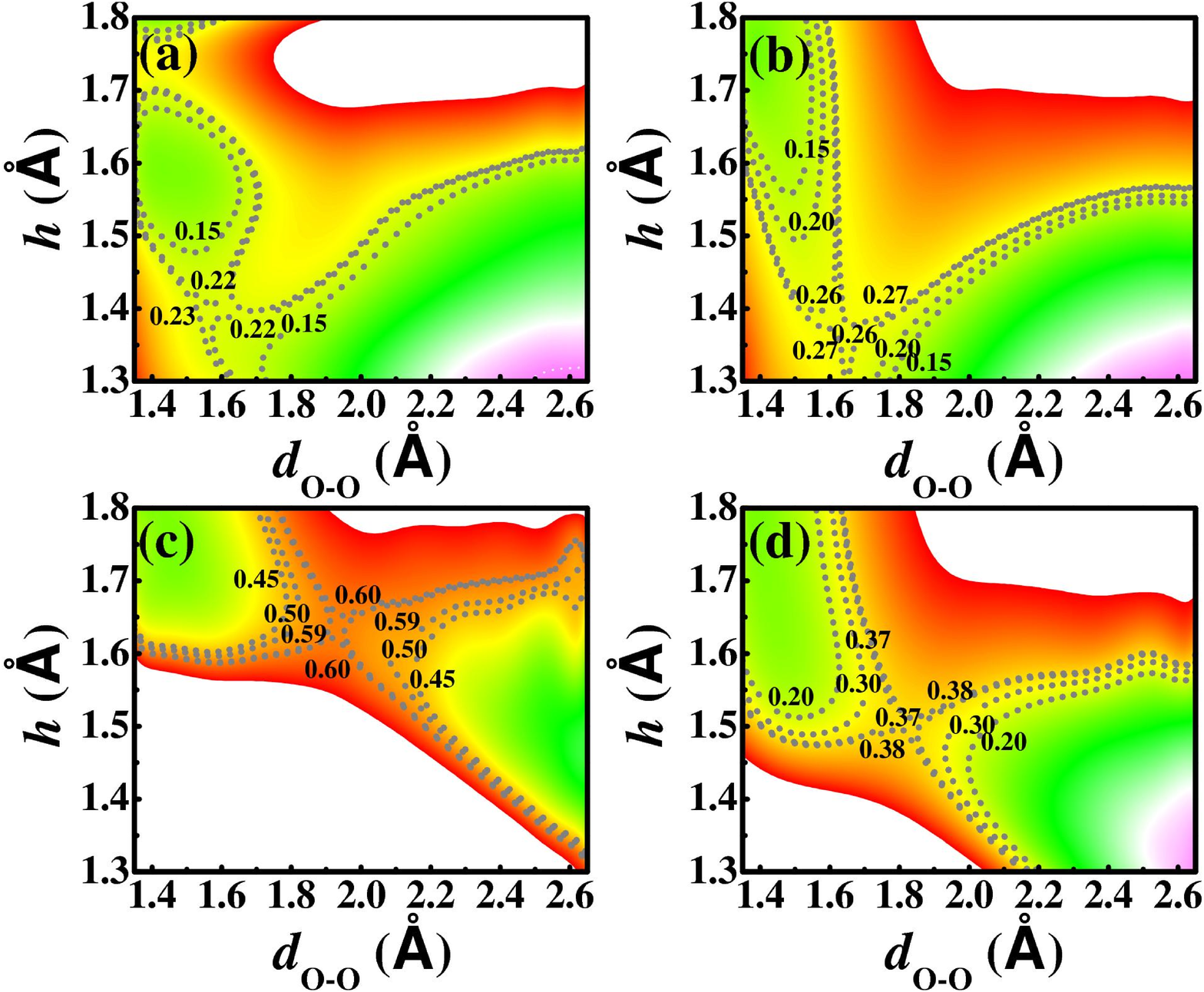}
\caption{\label{fig:fig7}}
\end{figure}
\clearpage
\begin{figure}
\includegraphics[width=1.0\textwidth]{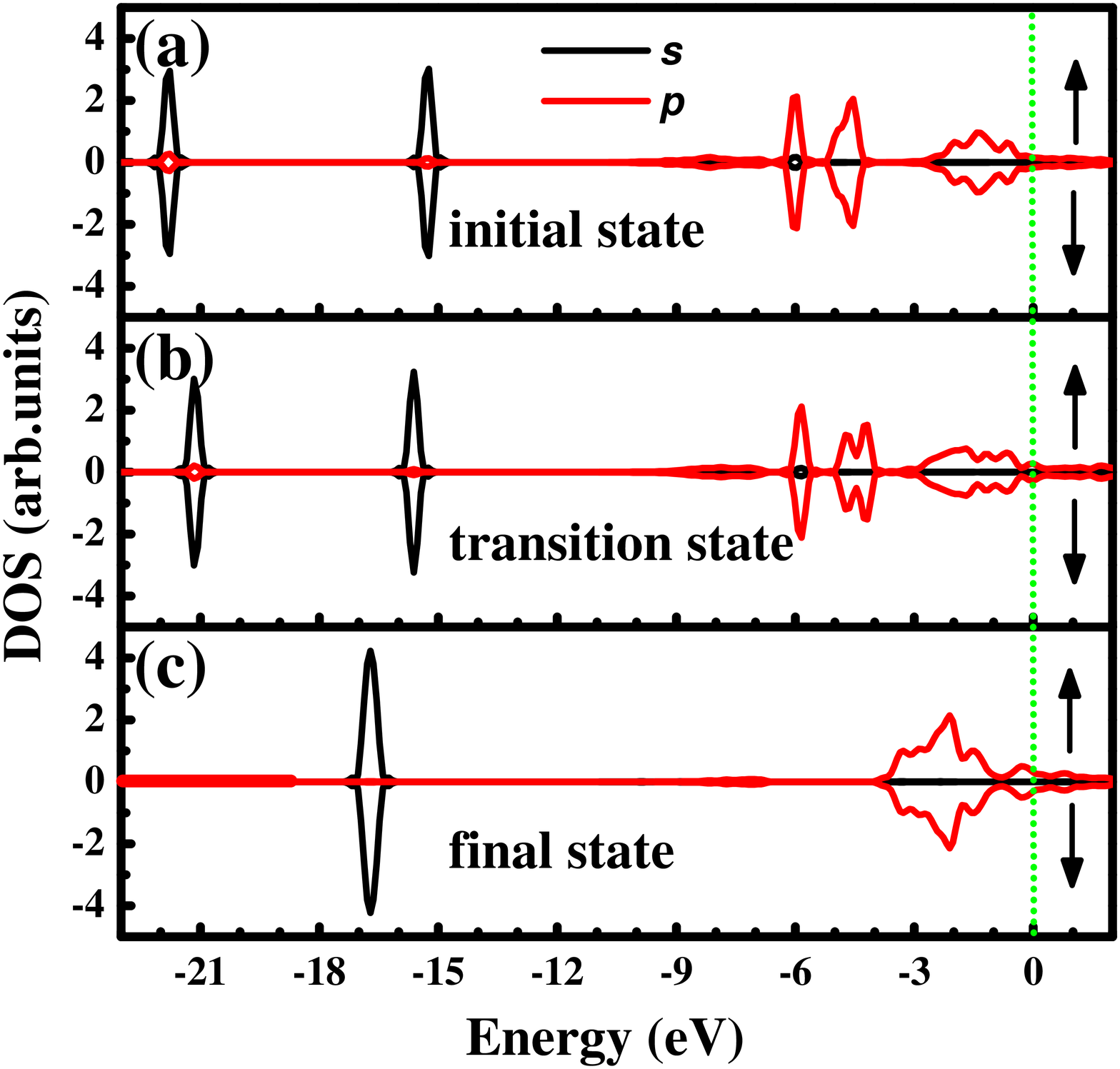}
\caption{\label{fig:fig8}}
\end{figure}
\clearpage
\begin{figure}
\includegraphics[width=1.0\textwidth]{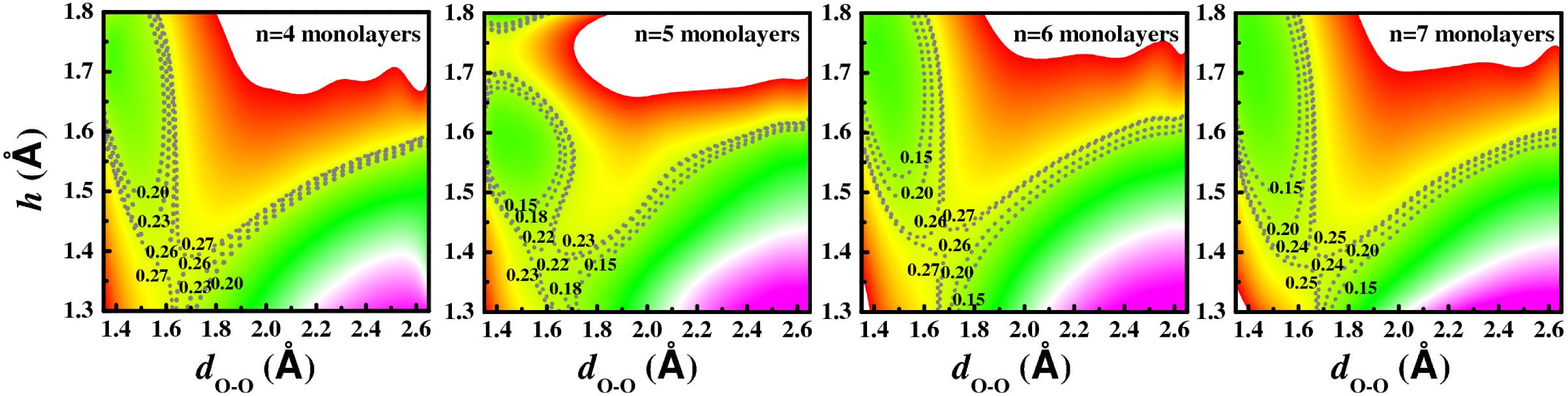}
\caption{\label{fig:fig9}}
\end{figure}

\end{document}